\documentclass[]{spieExo}  
 \usepackage{graphicx}
 \usepackage{amssymb}
 \usepackage{epstopdf}
\def\note#1{\relax} 

\title{Externally Dispersed Interferometry for Precision Radial Velocimetry} 

\author{D. J. Erskine\supit{a}, M. W. Muterspaugh\supit{b}\supit{,c}, J. Edelstein\supit{b}, J. Lloyd\supit{d},\\T. Herter\supit{d}, W. M. Feuerstein\supit{b}, P. Muirhead\supit{d}, and E. Wishnow\supit{b}
\skiplinehalf
\supit{a}{\normalsize Lawrence Livermore Nat. Lab., 7000 East Ave, Livermore, CA 94550 }\\
\supit{b}{\normalsize Space Sciences Lab. at Univ. of Calif., Berkeley, CA 94720-7450}\\
\supit{c}{\normalsize Townes fellow}\\
\supit{d}{\normalsize Astronomy Dept., Cornell University, Ithaca, NY 14853}}

\authorinfo{Further information: http://www.spectralfringe.org/EDI
\\D.E.:  erskine1@llnl.gov;
M.M: matthew1@ssl.berkeley.edu;
 J.E.: jerrye@ssl.berkeley.edu;
J.L.:  jpl@astro.cornell.edu;}

\begin{document}
 \maketitle
 
\large  
\begin{abstract}
Externally Dispersed Interferometry (EDI) is the series combination of a 
fixed-delay field-widened Michelson interferometer with a dispersive 
spectrograph.  This combination boosts the spectrograph performance for both 
Doppler velocimetry and high resolution spectroscopy.  The interferometer 
creates a periodic spectral comb that multiplies against the input spectrum 
to create moir\'e fringes, which are recorded in combination with the regular 
spectrum.  The moir\'e pattern shifts in phase in response to a Doppler 
shift.  Moir\'e patterns are broader than the underlying spectral 
features and more easily survive spectrograph blurring and common 
distortions.  Thus, the EDI technique allows lower resolution spectrographs 
having relaxed optical tolerances (and therefore higher throughput) to return 
high precision velocity measurements, which otherwise would be imprecise for 
the spectrograph alone.
\end{abstract}


\section{Introduction}

Externally Dispersed Interferometry (EDI) is a new technology for precision 
Doppler velocimetry that is both more efficient and less expensive than 
traditional high resolution spectrographs \cite{ErskinePatSuperimpose,ErskinePatEDI2002,ErskineEDITheory2003,SPIEscot,HiResSPIE,UVDiegoSPIE,TediOrlandoSPIE,ResBoostApJ2003,TediDiegoSPIE,OrlandoNoiseSPIE,ErskineGe2000,G.E.R.2002,GeVirgoApJ2006,Ge2002,Ge2003,Ge51Peg,VanEykenAAS2006}.  This technique 
increases the 
responsivity of a low-dispersion spectrograph to sharp spectral features 
which would normally be unresolved.  The first exoplanet results are arriving from these instruments as they are starting to come online.  
An EDI recently discovered a new exoplanet around the star HD 102195 
in Virgo\cite{GeVirgoApJ2006}. A multi-object EDI is being tested 
at the 2.5 m Sloan telescope\cite{VanEykenAAS2006}.  An NSF funded project 
is underway to field an EDI at the cassegrain output of the Mt.~Palomar 
Observatory 200 inch telescope in series with the TripleSpec near infrared 
(0.8--2.4 $\mu m$) spectrograph being built by Cornell University \cite{TripleSpecSPIE}.  This will find low mass exoplanets around low mass 
stars\cite{TediDiegoSPIE,TediOrlandoSPIE,OrlandoNoiseSPIE}.

An EDI is the series combination of a fixed-delay interferometer with a 
dispersive spectrograph (Fig.~\ref{fig:f1}).  The interferometer generates 
fringes which shift in phase in response to a Doppler velocity, and the 
disperser separates the fringes of different wavelengths on the detector 
array so that they can be detected at high visibility.  The advantages of an 
EDI over high-resolution spectroscopic velocimetry are three-fold:

\begin{figure}[!tbh]
\center{\includegraphics[height=7cm]{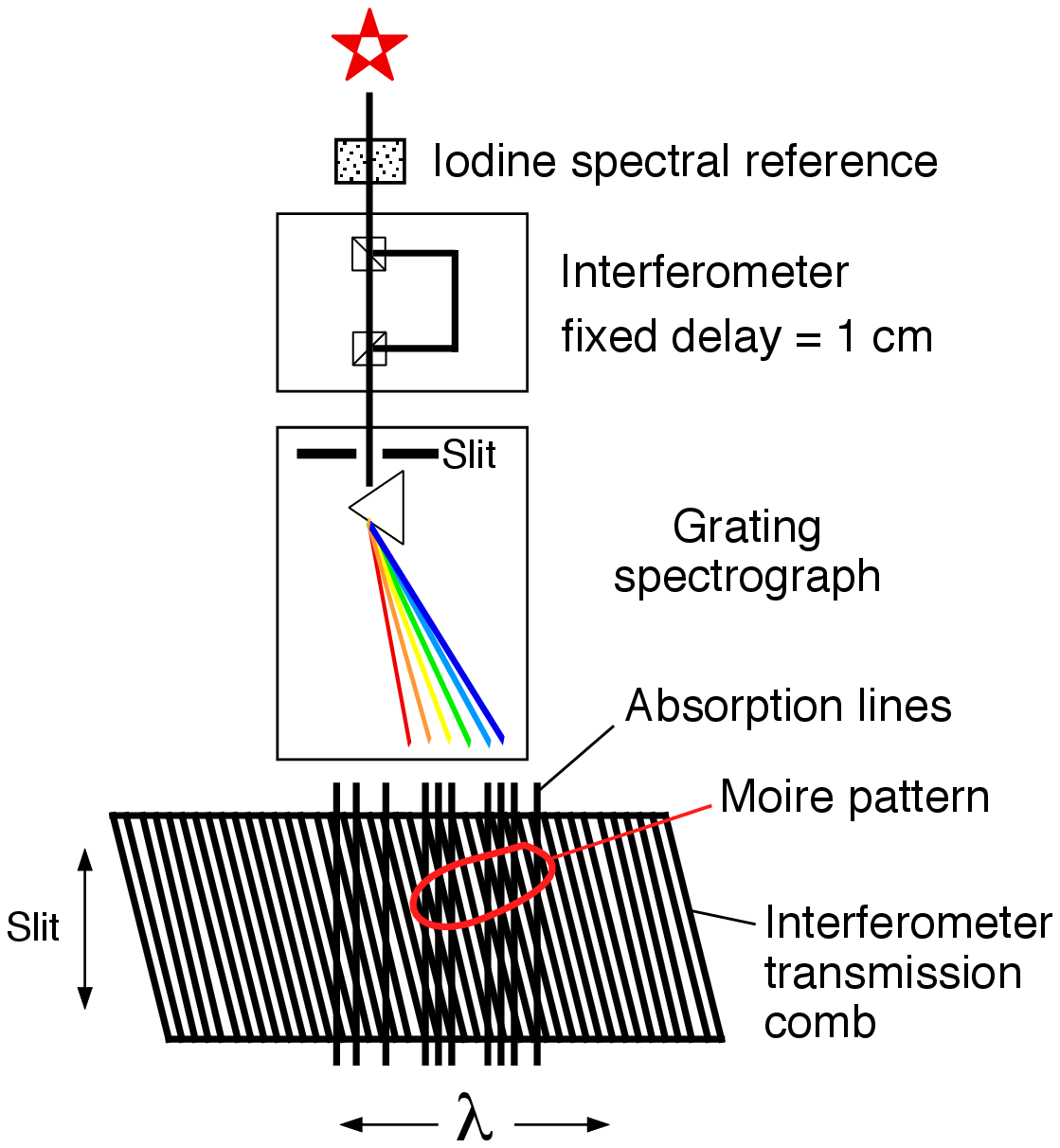}
\includegraphics[width=7.5cm]{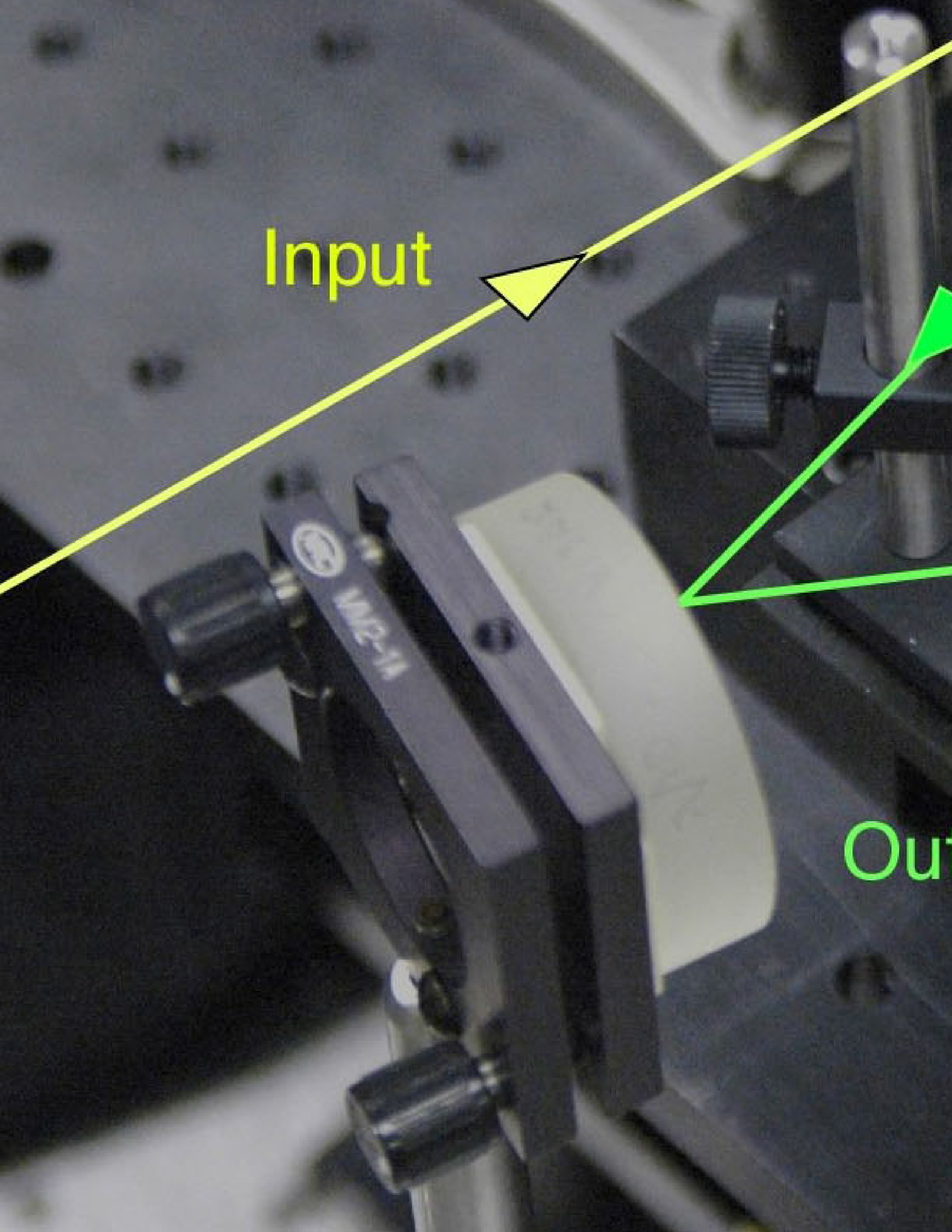}}
\caption[Fig1] {
[Left] EDI concept.  A sinusoidal transmission comb created by the interferometer is 
multiplied against the stellar spectrum, adding moir\'e fringes to the ordinary 
spectrum.  The moir\'e and ordinary components are separated during analysis.  
The moir\'e fringes change phase in proportion to the star's Doppler velocity. 
[Right] Example EDI apparatus, having dual outputs for high efficiency. 
\label{fig:f1}
\label{fig:f2}}
\end{figure}

\begin{enumerate}
\item {\bf The cost, weight, and size of the spectrograph and detector are 
greatly reduced.}  Much lower resolution spectrographs can be used to return 
precision Doppler velocities.  For example, the EDI used to detect the planet 
around HD 102195 had a resolution\cite{GeVirgoApJ2006} of only 
$\lambda / \Delta \lambda = 5000$.
\item {\bf The throughput and sensitivity of an EDI is increased.}  Lower 
resolution spectrographs usually have better throughput; the EDI preserves 
this property.  For example, the Exoplanet Tracker on the 2.1 meter Kitt Peak 
telescope can perform precision radial velocity (RV) measurements on sources 
as faint as 10th magnitude (15 minutes to 
achieve ~8 m/s for V=8, Ref.~\citenum{ETnoaoWebpage2006}); this compares favorably to 
HIRES on the 10 meter Keck telescopes!  Wider bandwidths can be used using the interferometer signal as a 
wavelength fiducial linking spectral features across the band.  The increased 
number of spectral lines covered improves RV precision and sensitivity.  
Smaller telescopes can be used for a given target brightness, which often 
have more time available for higher cadence observations.  Alternatively, more 
targets are accessible to large telescopes, possibly including the faint stars 
targeted for deep transit searches, including the Kepler field.
\item {\bf The EDI measurement is more robust to systematic and instrumental 
effects.}  The stepping manner of measuring moir\'e pattern phases eliminates 
fixed pattern errors such as those due to detector or optics blemishes by 
acting as a built-in flatfielding---they are not synchronous with the 
stepping.  Common spectrograph irregularities, such as drifts in the 
focal spot position and diameter on the detecting array, do not strongly 
affect the EDI observable of fringe phase.  This is due to the differential 
nature of a moir\'e pattern in comparing the stellar and interferometer 
sinusoidal combs.  The interferometer comb can be thought of as a set of 
fiducials coupled to the science signal and follow it through the 
instrument.  Distortions affecting the science signal are also applied to the 
interferometer comb.  We have estimated that the EDI is 1000 to 10,000 times 
more robust to translation of the focal spot, and 10 to 100 times more robust 
to change in focal spot diameter.
\end{enumerate}

\section{Instrument method}

The transmission spectrum of an interferometer of delay $\tau$ 
is sinusoidal versus wavenumber ($\nu = 1/\lambda$).  Passing starlight 
through an interferometer multiplies the periodic sinusoidal interferometer 
comb by the stellar spectrum.  Heterodyning between similar spacing of 
features (along the dispersion axis) in the stellar spectrum and the 
sinusoidal comb creates beating between frequencies.  These ``Moir\'e 
patterns'' are broader and more easily survive the blurring of a spectrograph 
than the original narrow features.

\begin{figure}[!tbh]
\center{\includegraphics[height=5cm]{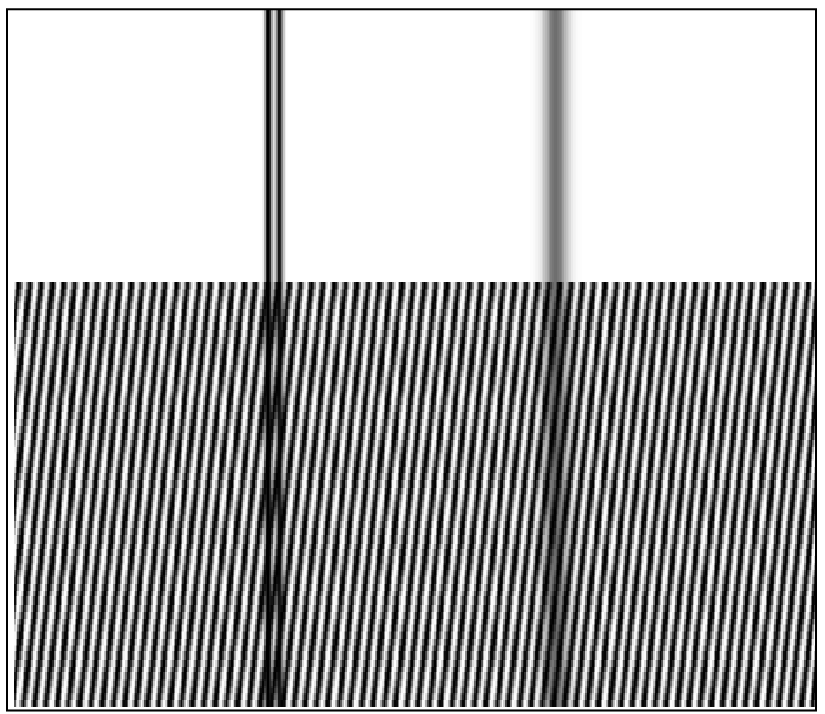}
\includegraphics[height=5cm]{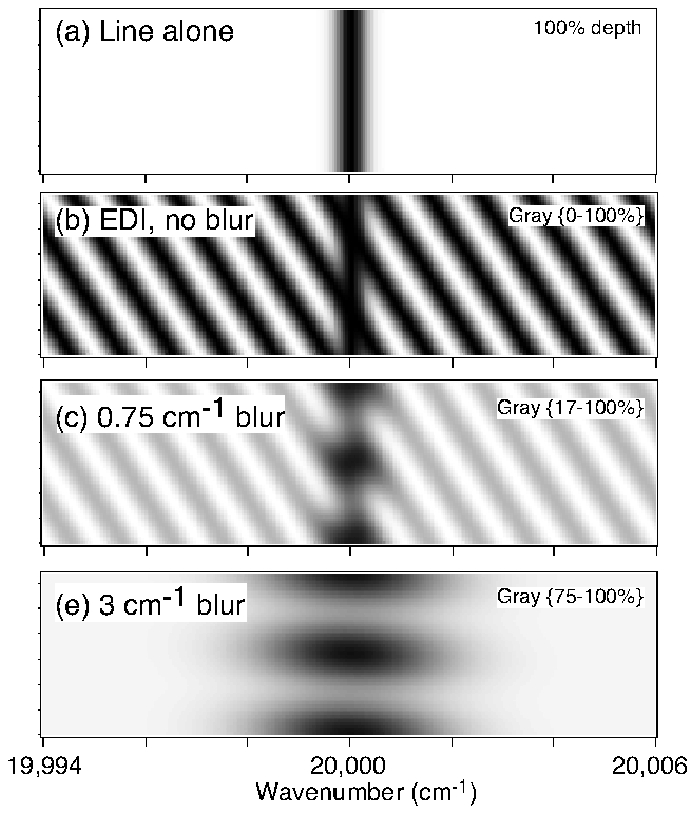}}
\caption[Fig3] {
(Left) The moir\'e fringes created by overlaying a sinusoidal transmission 
on an input spectrum distinguishes a doublet from a singlet, satisfying the 
classical definition of a resolution increase.
(Right) Moir\'e fringes persist under spectrograph blurring.  
\label{fig:f3}}
\end{figure}

\begin{figure}[!tbh]
\center{\includegraphics[width=8cm]{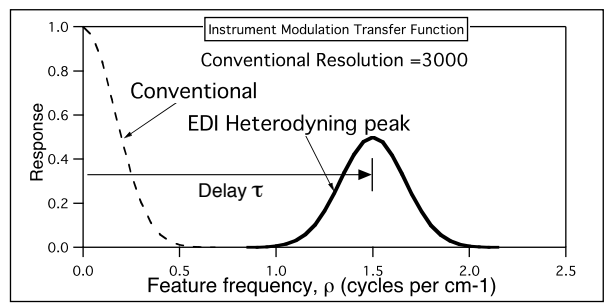}}
\caption[Fig7] {
The EDI sensitivity has a peak centered at high spatial frequencies chosen by 
the interferometer delay $\tau$.  This can be placed where more Doppler 
information content exists in the stellar spectrum, which is often at high 
spatial frequencies beyond the sensitivity peak of the spectrograph alone 
(dashed curve).
\label{fig:f7}}
\end{figure}

A Doppler velocity change in the stellar spectrum produces a phase change in 
the moir\'e phase.  However, a slight drift in the interferometer delay, such 
as due to thermal changes, also will cause a phase shift.  The solution is to 
simultaneously record a calibrant spectrum (such as iodine vapor, for visible 
light) along with the stellar spectrum.  The Doppler shift is the difference 
between the stellar and calibrant moir\'e phases.
 
\section{Doppler quality factor}

\begin{figure}[!tbh]
\center{\includegraphics[width=8cm]{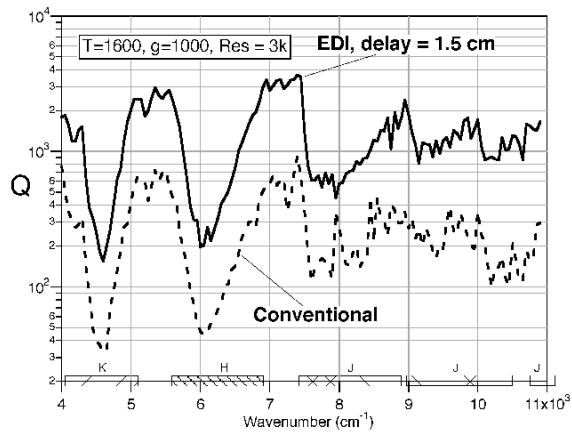}}
\caption[Fig8] {
A spectrum's Doppler quality ($Q$) versus wavenumber ($\nu$), calculated for 
a stellar model having temperature T=1600, gravity g=1000, and no rotational 
blurring.  Shown are conventional $Q$ and EDI $Q$ for native spectrograph 
resolution Res=3k and interferometer delay of 1.5 cm. 
\label{fig:f8}}
\end{figure}

The radial velocity measurement noise is a function of the characteristic 
spectral derivative and the number of photons recorded. The photon limited 
velocity noise ($\delta V$) in a radial velocity measurement is given by \cite{Connes1985}
\begin{equation}
\delta V={c\over Q}{1 \over \sqrt{N}} 
\label{eq:11} 
\end{equation} 
where $N$ is the total number of detected photons summed over the bandwidth 
in question. The $Q$ is a dimensionless normalized RMS average of the 
spectrum's derivative. Therefore, Q is high and the noise is low when the 
spectral lines are numerous and narrow. The Doppler velocity signal to noise 
ratio is proportional to $Q$.  Reference~\citenum{OrlandoNoiseSPIE} shows how to calculate 
Q for a stellar spectrum of a cool star (Fig.~\ref{fig:f8}). 
For calculating $\delta V$ when $Q(\nu)$ or the intensity $dN/d\nu$ varies 
with $\nu$, one integrates $Q^{2}(\nu) dN/d\nu$ over the bandwidth in 
question, takes the square root, and substitutes that value for $Q \sqrt{N}$ 
in Eq.~\ref{eq:11}. 

\section{Comparison to other methods}

The EDI is distinguished from {\it internally} dispersed interferometers such 
as the Spatial Heterodyning Spectrometer\cite{Harlander1992} by having the 
dispersive element {\it outside} the interferometer cavity rather than 
inside, so that ray angles inside the interferometer, which affects fringe 
phase, are the same for all wavelengths.  This allows the simultaneous 
bandpass for the EDI to be essentially unlimited, (whereas for the SHS it is 
very limited).

In EDI Both complementary interferometer outputs can be directed to the spectrograph 
(Fig.~\ref{fig:f2} right), so for ideal optics every photon that enters the 
interferometer passes to the spectrograph and contributes to a Doppler 
velocity signal.  


In conventional Doppler radial velocimetry, a dispersive spectrograph is used 
having a minimum resolving power of about 50,000.  Resolutions less than this 
are insufficient to prevent neighboring spectral features from blending 
together, reducing the slope on the edge of each line.  (A large slope is 
needed to produce a detectable intensity change for a given Doppler 
wavelength change.)  Unfortunately, spectrographs having this minimal 
resolution are very large and expensive for large telescopes, because 
spectrograph size scales with resolution, bandwidth and telescope diameter, 
and spectrograph cost and weight scale nonlinearly with spectrograph size.


\begin{figure}[!tbh]
\center{\includegraphics[width=2.7in]{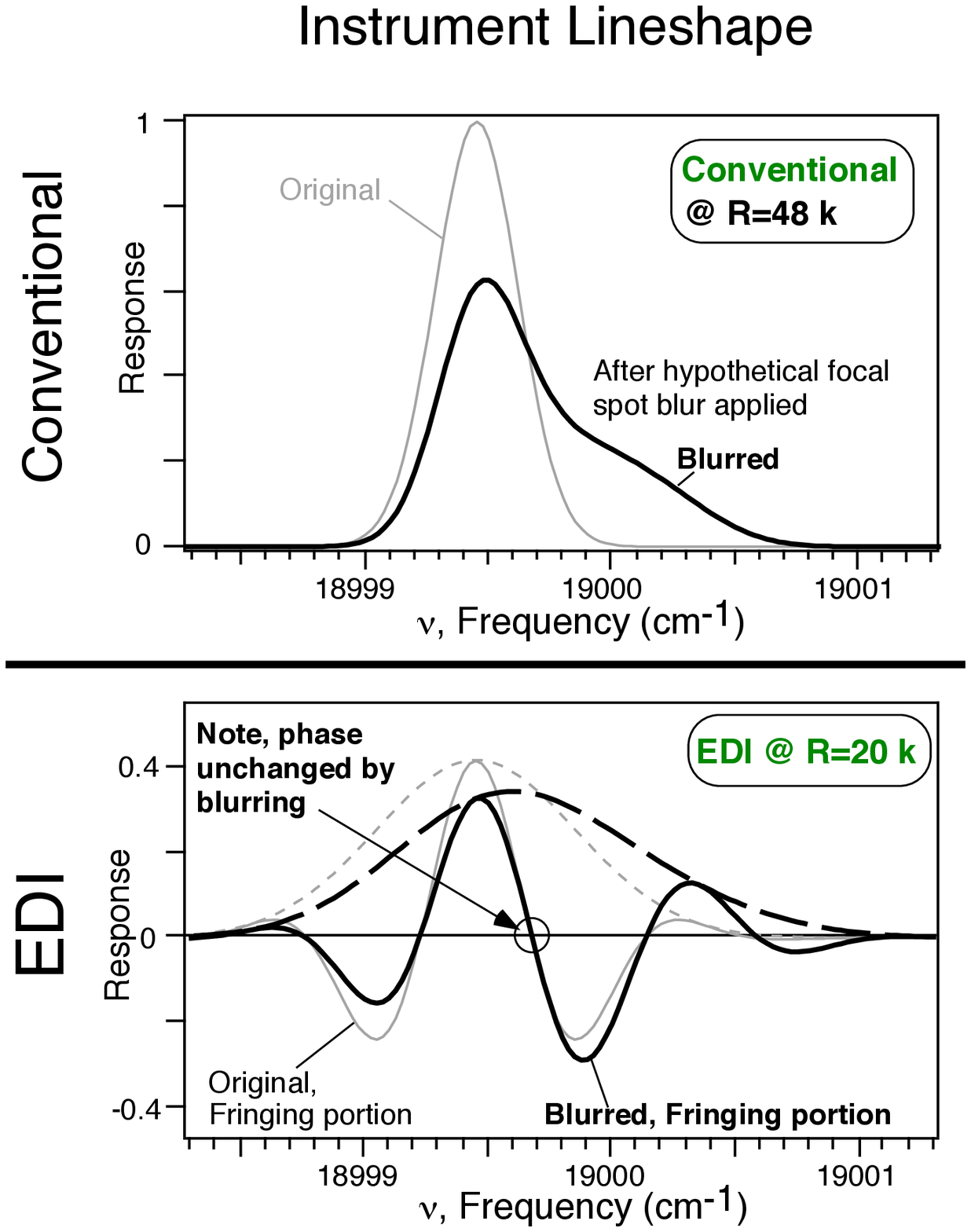}
\includegraphics[width=3.in]{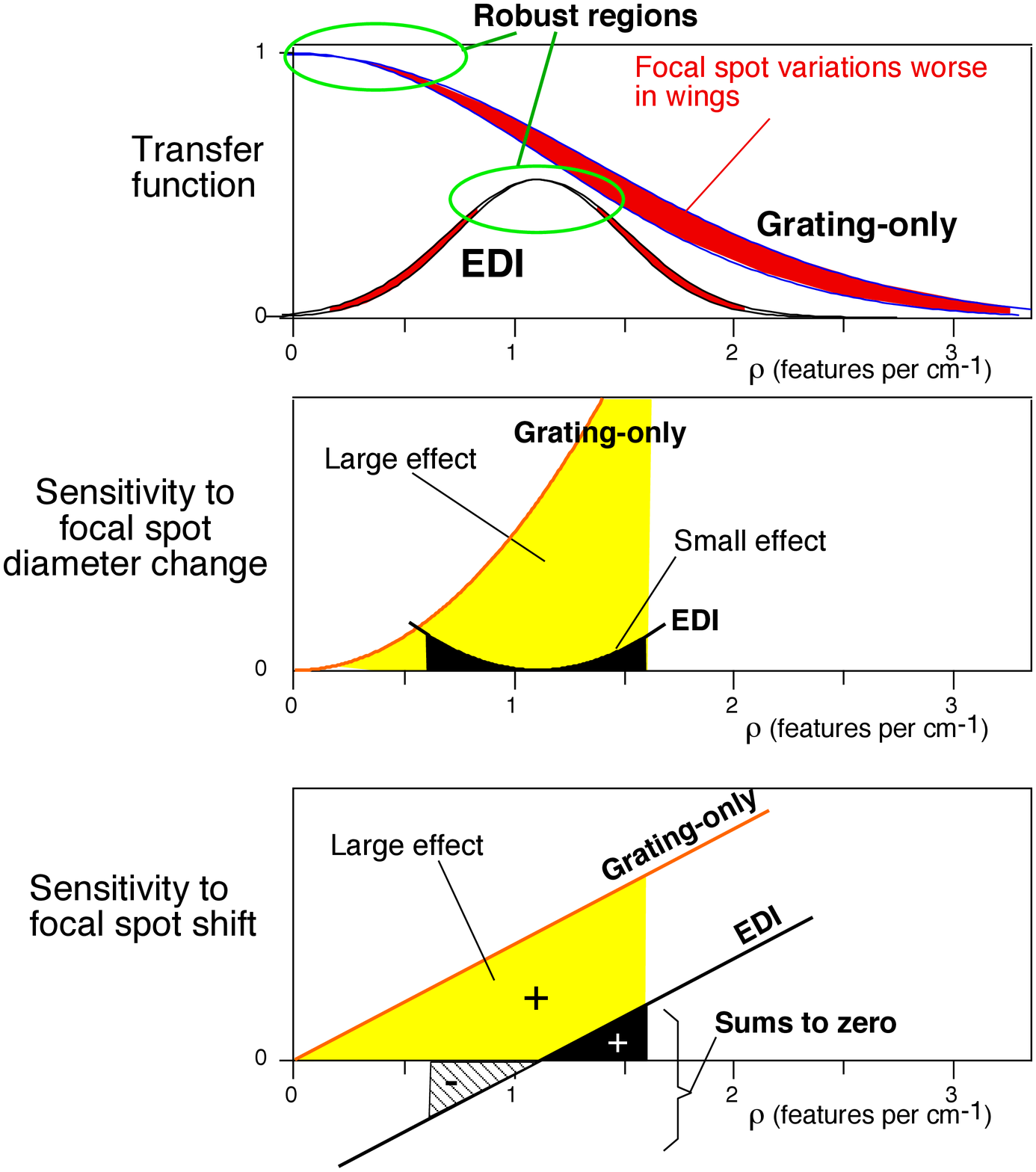}}
\caption[Fig5] {
(Left) Numerical simulation of the effect of spectrograph instrument line 
shape distortion (black vs gray) on conventional (top) and EDI data 
(bottom).  Note that the fringe phase zero crossing (circle) is unchanged.
(Right) Robustness of EDI instrument line shape to focal spot motion and 
diameter change.
\label{fig:f5}}
\end{figure}

\section{Infrared RV and T-EDI}

RV Doppler measurements\cite{Mar:98::} have proved to be 
the most successful way of finding extrasolar planets to date.  It appears 
this method may fall short of actually finding Earthlike planets in the 
habitable zones of Sunlike stars, because the 0.1 ${\rm m\,s^{-1}}$ velocity 
amplitude induced on the star around the system's barycenter is smaller than 
pulsations of the stars themselves.

\begin{figure}[!tbh]
\center{\includegraphics[width=2.7in]{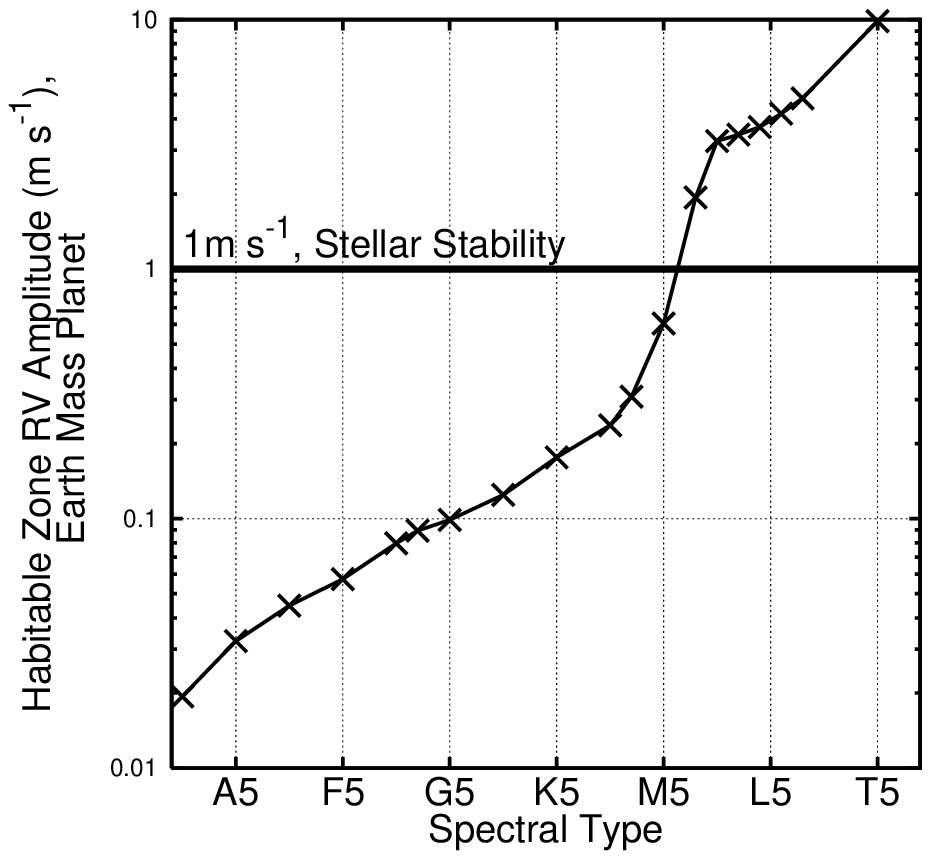}
\includegraphics[width=2.7in]{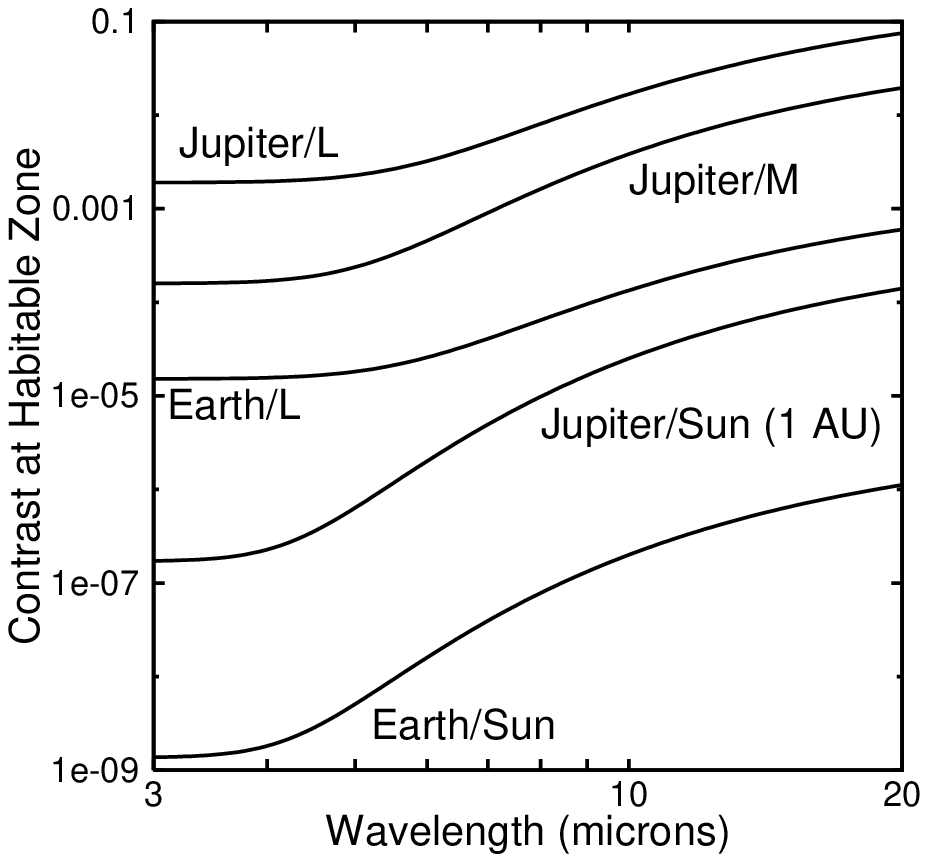}}
\caption[Fig0] {
(Left)  The radial velocity signature of an earth-mass planet in the 
habitable zone as a function of stellar spectral type.  The RV technique can 
detect Earthlike planets around stars later than $\sim {\rm M5}$; the RV 
signature for earlier spectral types is often smaller than the RV stability 
of the star's atmosphere.  
(Right)  The star-planet contrast for planets in the habitable zones of a 
variety of dwarf star types, as a function of observing wavelength.  
Blackbody spectra are assumed for both star and planet, where the planet also 
reflects starlight.  Not shown are spectral features in certain windows which 
might decrease the contrast.
\label{fig:f0}}
\end{figure}

RV searches targeting lower mass stars present an opportunity to overcome this 
limitation because two factors increase the velocity signatures of habitable exoplanets; see Figure \ref{fig:f0}.  First, the star's reflex 
motion velocity is inversely proportional to the star's mass; this presents 
one factor of 10 improvement.  Second, low-mass stars have lower surface 
temperatures than the Sun, and their habitable zones are correspondingly 
closer.  RV amplitude is inversely proportional to the square root of the 
star-planet separation; this adds another factor of $\sim 3$ improvement.

Finding planets around low mass, cool stars is also of interest for direct 
imaging studies because the contrast requirements are lowered.  This can 
ease design requirements by a factor of $\sim 1000$.

Low mass stars are intrinsically faint and most of their light is found at 
near infrared wavelengths.  No current system is capable of observing these 
stars with RV precisions better than $\sim 100$ ${\rm m\,s^{-1}}$; a few 
early M dwarfs are accessible to Iodine-cell high resolution RV, but these 
extend only to about M3.

The TripleSpec instrument is a low resolution ($R \sim 2700$), wide 
bandwidth (simultaneous coverage of $800-2400\,{\rm nm}$) spectrograph being built for the Palomar 200'' telescope by Cornell, for 
which an EDI unit is being constructed, (hence ``T-EDI'') .  This EDI combination boosts the TripleSpec effective 
spectral resolution by a factor of $3-7\times$, and the associated Doppler 
velocity precision, while maintaining the high throughput and bandwidth of 
the low resolution system.
The large bandwidth and high throughput of the resulting instrument will allow 
faint, cold stars to be studied.  
It will be capable of Doppler velocity
measurements with precisions of better than 3 ${\rm m\,s^{-1}}$, 
and for 
very faint stars (infrared $H\sim 13$ magnitudes) 10 ${\rm m\,s^{-1}}$.
This will be used to 
search for Earthlike planets in the habitable zones of low mass stars.

\acknowledgments     
\normalsize
Thanks to Didier Saumon, Mark Marley and Richard S. Freedman for high resolution stellar models. This work was performed with support from the National Science Foundation (awards AST-0504874 \& AST-0505366), and under the auspices of the U.S. Department of Energy by the University of
California, Lawrence Livermore National Laboratory  under contract No. 
W-7405-Eng-48.

\normalsize
\bibliography{Mine3,OthersAstro4,misc2006}   
\bibliographystyle{spiebib}   

 \end{document}